\documentclass{article}

\newlength\titlebox 

\usepackage[dvips]{graphicx}

\usepackage{amssymb,amsfonts,amsmath,url}

\begin{document}
\title{Identifying network communities with a high resolution}

\author{Jianhua Ruan$^1$ and Weixiong Zhang$^{1,2}$\\
Department of Computer Science$^1$ and Department of Genetics$^2$\\
Washington University in St. Louis, St. Louis, MO 63130, USA\\
jruan@cse.wustl.edu, zhang@cse.wustl.edu
}

\maketitle

\begin{abstract} Community structure is an important property of complex networks. An automatic discovery of such structure is a fundamental task in many disciplines, including sociology, biology, engineering, and computer science. Recently, several community discovery algorithms have been proposed based on the optimization of a quantity called modularity (\emph{Q}). However, the problem of modularity optimization is NP-hard, and the existing approaches often suffer from prohibitively long running time or poor quality. Furthermore, it has been recently pointed out that algorithms based on optimizing \emph{Q} will have a resolution limit, i.e., communities below a certain scale may not be detected. In this research, we first propose an efficient heuristic algorithm, \emph{Qcut}, which combines spectral graph partitioning and local search to optimize \emph{Q}. Using both synthetic and real networks, we show that \emph{Qcut} can find higher modularities and is more scalable than the existing algorithms. Furthermore, using \emph{Qcut} as an essential component, we propose a recursive algorithm, \emph{HQcut}, to solve the resolution limit problem. We show that \emph{HQcut} can successfully detect communities at a much finer scale and with a higher accuracy than the existing algorithms. Finally, we apply \emph{Qcut} and \emph{HQcut} to study a protein-protein interaction network, and show that the combination of the two algorithms can reveal interesting biological results that may be otherwise undetectable.

\noindent Supplemental file: \url{http://cic.cs.wustl.edu/qcut/supplemental.pdf} .


\end{abstract}

\section{Introduction}

Many complex systems can be represented as networks, where vertices are the elements in a system, and edges represent relationships between pairs of elements. Examples include social networks~\cite{NewCol2004}, genetic networks~\cite{Bad2002}, and the Internet~\cite{Kle2001}. Much effort has been devoted to the study of topological properties that are common to many networks, such as the small-world property, power-law degree distributions, and high clustering coefficients~\cite{Alb2002,NewmanSIAM03}. 

Another important property of complex networks that has drawn a great deal of attention recently is the so-called community structure, i.e. the existence of some natural division of a network such that the vertices in each sub-network are highly associated among themselves, while having relatively fewer/weaker connections with the rest of the network~\cite{Newman2006,New2006Eig}. Because communities are relatively independent of one another structurally, it is believed that each of them may correspond to some fundamental functional unit. For example, a community in genetic networks often contains genes with similar functions, and a community on the World Wide Web may correspond to web pages related to similar topics. Identifying and analyzing such communities from a large network, therefore, provides a functional dissection of the network, and sheds light on its organizational principles. Furthermore, community structures may provide key insights into some uncharacterized properties of a system. For example, attempts have been made to identify and characterize communities (or called functional modules sometimes) in biological networks, leading to \emph{in silicon} predictions of the functions of some genes~\cite{Per2004,Spi2003,Wil2004}.

Community discovery is similar but not equivalent to the conventional graph partitioning problem~\cite{graphPart}, both of which require  clustering vertices into groups~\cite{Newman2006}. In a conventional graph partitioning, the graph is assumed to be always partitionable, and the number of partitions is usually predefined. The challenges in community discovery, however, are two-fold: (1) what constitutes a community, and (2) how to effectively find such communities. Although several definitions of communities have been proposed, none has been universally accepted~\cite{Fla2002,Rad2004}. The general agreement is that a community discovery algorithm needs to decide by itself the most appropriate community structure without prior knowledge about a network, and should be able to distinguish between networks having good community structures and networks with essentially random structures.

Instead of explicitly defining communities, Newman and Girvan recently proposed a quantitative measure, called modularity ($Q$), to assess the quality of a community structure, and formulated community discovery as an optimization problem~\cite{NG2004}. The idea has since been widely adopted, and several algorithms have been developed to optimize $Q$, with good performance in practice~\cite{AAAI06, WhiteS05, Newman2006,Danon2005,Gui2005,CNM2004}. However, it has been shown that optimizing $Q$ is NP-hard~\cite{brandes-2006}, which means an efficient optimal algorithm for the problem is unlikely to exist. The fastest algorithm available uses a greedy strategy and suffers from poor quality~\cite{CNM2004}. A more accurate method is based on simulated annealing, which has a prohibitively long running time for large networks~\cite{Gui2005}. The best existing algorithm in terms of both efficiency and effectiveness is due to Newman~\cite{Newman2006}.

On the other hand, although empirical studies have shown that modularity optimization is often an effective way to detect communities, several researchers and ourselves have observed that this strategy may lead to a resolution limit problem~\cite{For2007,Muf2005}. Briefly, by optimizing modularity, communities that are smaller than a certain scale or have relatively high inter-community connectivities may be merged into a single community. This limit, therefore, has cast some doubts on the effectiveness of modularity optimization for community discovery~\cite{For2007}.

In this paper, we first present an efficient heuristic algorithm, called \textit{Qcut}, to optimize $Q$ by combining spectral graph partitioning and local search. We show that the algorithm is able to find higher $Q$ values and is more scalable to large networks than the best existing algorithms. For synthetic networks without the resolution limit problem, we also show that \emph{Qcut} can achieve a much higher accuracy than the existing algorithms in recovering the known communities. 

More importantly, we show that, although modularity optimization has a resolution limit, it is effective in detecting communities at a coarse-grained level, i.e. vertices belonging to the same community tends to be grouped together. This observation is the key for our second algorithm, called \emph{HQcut}, to solve the resolution limit problem. The \emph{HQcut} algorithm recursively applies \emph{Qcut} to divide a community into sub-communities. In order to avoid over-partitioning, we use a statistical test to determine whether a community indeed contains intrinsic sub-community. We demonstrate the effectiveness of \emph{HQcut} on a number of synthetic and real networks, and show that \emph{HQcut} can successfully detect communities at a much finer scale and with a higher accuracy than the algorithms based on modularity optimization alone. 

Furthermore, we discuss two primary causes of the resolution limit problem in practice. First, real-world networks often have diverse community sizes. Some small communities may accidentally connect to one another by a few edges due to noises. Second, real-world networks may have hierarchical community structures, i.e. a community may contain several relatively highly interconnected sub-communities. It is crucial to be able to discern such subtle community structures. Therefore, we propose a statistical test to differentiate the two cases, and show some interesting statistic in real-world networks.

Finally, we apply \emph{Qcut} and \emph{HQcut} to study a protein-protein interaction network in the budding yeast, and analyze the biological significance of the resulting communities. We show that combining the results of these two algorithms can reveal some interesting biological results that may be otherwise undetected.

\section{Community identification by modularity optimization}

Given a network with $N$ vertices and $M$ edges, and a partition that divides the vertices into $k$ communities, the modularity function is defined as 
\begin{equation}
Q = \sum_{i=1}^{k} \left(\frac{e_{ii}}{M} - \left(\frac{a_i}{2M}\right)^2\right),
\label{eqn:Q}
\end{equation}
where $e_{ii}$ is the number of edges within community $i$, and $a_{i}$ is the total degree for the vertices in community $i$~\cite{NG2004}. 
The $Q$ function measures the fraction of edges falling within communities, subtracted by what would be expected if the edges were randomly placed. A larger $Q$ value indicates stronger community structures. If a partition gives no more intra-community edges than would be expected by chance, $Q \le 0$. For a trivial partition with a single cluster, $Q = 0$. Given the definition of $Q$, the community discovery problem is to find a partition of the network that optimizes $Q$.

\subsection{The Qcut algorithm}
Since the optimization of $Q$ is NP-hard, we adopt a two-stage procedure, namely, partitioning and refining. In the partitioning stage, a spectral graph partitioning algorithm is applied recursively to divide a network until no improvement of $Q$ can be achieved. This step provides an efficient approximate solution that gives a reasonably good $Q$ value. The spectral algorithm is described in the supplemental file.

In the refining stage, a local search strategy is applied to improve $Q$ as much as possible. We repeatedly consider the following three types of operations: (1) \emph{migration}: move a vertex from its current community to another one; (2) \emph{merge}: combine two communities to form a single one; and (3) \emph{split}: divide a community into two smaller ones. In this process, we use the steepest ascent hill climbing heuristic, i.e., the algorithm always executes the operation that gives rise to the highest $Q$. 

It is much more expensive to search for a good split than for a migration or merge. Therefore, we consider split only if no migration and merge can improve $Q$. We use the same spectral graph partitioning algorithm used in the first stage to suggest possible split operations.

To efficiently identify a good migration or merge operation, we pre-compute the change to $Q$ for each potential migration or merge. The change to $Q$ incurred by moving vertex $v$ from its current community $i$ to a new community $ j$ can be computed by:
\begin{equation}
\Delta Q^{migration}(v, i, j) = \left\{ \begin{array}{ll}
\frac {d_j^v - d_i^v}{M} + \frac{d^v(a_i - a_j - d^v)}{2M^2} & \mbox{if $i \neq j$}; \\
%(d_j^v - d_i^v)/M + d^v(a_i - a_j - d^v)/2M^2 & \mbox{if $i \neq j$}; \\
0 & \mbox{if $i = j$}. \end{array} \right.
\label{eqn:migration}
\end{equation}
where $d_i^v$ and $d_j^v$ are the numbers of connections that $v$ has in communities $i$ and $j$, respectively, and $d^v$ is the total degree of vertex $v$. An intuitive interpretation of Equation~(\ref{eqn:migration}) is straightforward: in order to improve $Q$, we should try to move $v$ to a community that is relatively smaller (i.e., $a_i > a_j + d^v$), and where $v$ has more friends (i.e., $d_j^v > d_j^v$). Given an initial partition, we compute all $\Delta Q^{migration}$, and cache them in a table $T = (t_{vj})_{N \times K}$, where $N$ is the number of vertices, $K$ is the number of communities, and $t_{vj}$ is the potential change to $Q$ if we move vertex $v$ from its current community to community $j$. $T$ can be efficiently computed with matrix algebra. It may first seem that the table is a dense matrix, taking $O(NK)$ space to store and $O(NK)$ time to search. In fact, it can be shown that we do not need to compute $t_{vj}$ when $d_j^v = 0$, since the corresponding migration will not give the highest improvement of $Q$, even if it is positive (Supplemental file). Therefore, for a sparse network, most entries in $T$ can be set to zero, resulting in a sparse matrix.

Similarly, we can also compute the potential change to $Q$ if communities $i$ and $j$ are combined:
\begin{equation}
\Delta Q^{merge}(i, j) = \left\{ \begin{array}{ll}
\frac {e_{ij}}{M} - \frac{a_i  a_j}{2M^2} & \mbox{if $i \neq j$}; \\
%e_{ij}/M - a_ia_j/2M^2 & \mbox{if $i \neq j$}; \\
0 & \mbox{if $i = j$}. \end{array} \right.
\label{eqn:merge}
\end{equation}
where $e_{ij}$ is the number of edges connecting the two communities. Based on this equation, we can compute a table $S=(s_{ij})_{k \times k}$, where $s_{ij}$ is the potential change to $Q$ if communities $i$ and $j$ are merged.

Given $S$ and $T$, we then choose the operation that can result in the highest improvement to $Q$. This continues as long as there is some positive entry in $S$ or $T$. Importantly, after an operation is taken, we do not need to re-compute the entire tables, since most of the entries in $S$ and $T$ remain unchanged. 

As can be seen from Equations~(\ref{eqn:migration}) and~(\ref{eqn:merge}), each operation will improve $Q$ by at least $\frac{1}{M^2}$. Therefore, the algorithm will terminate in at most $M^2$ iterations, while in practice it usually terminates much sooner.

\subsection{A related algorithm}
Newman recently proposed a method that is also based on spectral graph partitioning and local search~\cite{Newman2006}. \emph{Qcut} significantly differs from the Newman's algorithm in two aspects. First, for the spectral partitioning, our algorithm utilizes the Laplacian matrix of a network, while his method deals with a so-called modularity matrix. It has been shown that both spectral partitioning methods can approximately optimize $Q$~\cite{Newman2006,WhiteS05}. Since the Laplacian matrix is typically sparse, while the modularity matrix is almost a complete matrix, our algorithm has a much lower memory requirement, and is more scalable to large networks. Second, the Newman's algorithm uses a Kernighan-Lin heuristic after each partitioning to switch members in two neighboring communities. Therefore, the refinement decision in his algorithm is only made locally. In contrast, a vertex can be moved to any communities in the refining stage of our algorithm, and therefore the decision is made globally. Furthermore, \emph{Qcut} also considers community merges and community splits to further improve $Q$.

\section{Limitation of modularity optimization and a solution}

Equation~(\ref{eqn:merge}) implies that in the final community structure identified by \emph{Qcut}, the number of edges, $e_{ij}$, connecting communities $i$ and $j$ has to be less than $a_ia_j/2M$, which can be interpreted as the expected number of edges connecting $i$ and $j$. If this condition is not satisfied, the algorithm will merge $i$ and $j$ in order to improve $Q$. This condition is intuitive: when two sub-networks are connected by a higher-than-expected number of edges, they are probably related, and therefore should not be partitioned into two communities. 

\begin{figure}
\center 
\includegraphics[scale=.5]{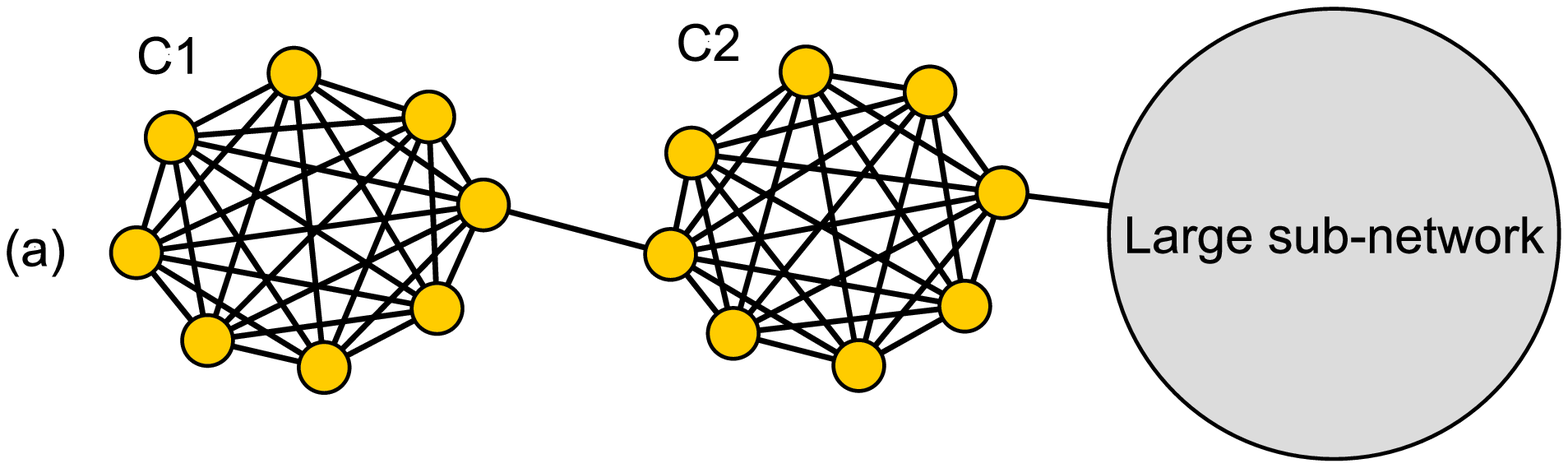}\\
\includegraphics[scale=.5]{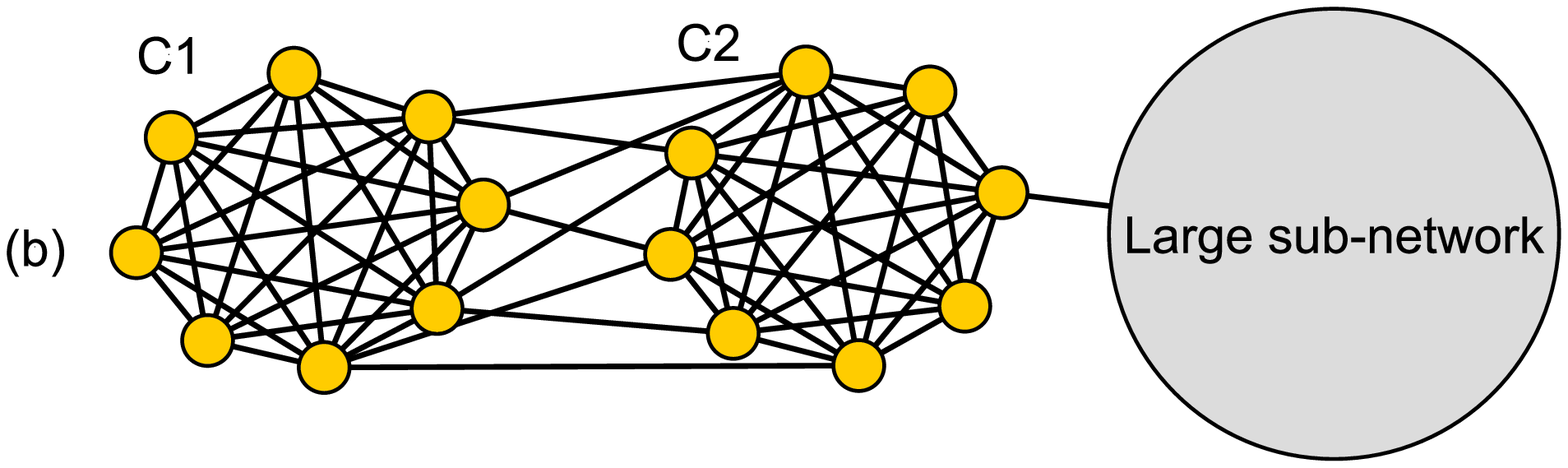}
\caption{{\label{fig:subcomm}} (a): two affiliated communities. (b): two associated communities.}
\end{figure}

However, consider the network in Fig.~\ref{fig:subcomm}(a), where two cliques are connected by a single edge. If there are no other vertices, the two cliques clearly form two communities. It becomes interesting, however, when one of the cliques is connected to a large network via a single edge. When the number of edges in the entire network, $M$, is greater than a threshold of $a_ia_j/2$, the expected number of edges between the two cliques, $a_ia_j/2M$, becomes smaller than one. Consequently, the two cliques will be considered as a single community, according to Equation~(\ref{eqn:merge}). The fact that modularity optimization cannot reveal communities that are smaller than a certain scale received attention lately in~\cite{For2007}, and was referred to as the resolution limit problem. 

The resolution limit has some significant impact in practice. Real-world networks often contain both large and small communities. In addition, many real-world networks such as social or biological networks are constructed from survey or experimental data, and therefore may contain errors. If two small communities are accidentally connected by a false edge, they will be non-separable by modularity optimization. The limitation, therefore, is partially due to the assumption that all edges in a network are reliable.

Furthermore, the modularity function is also limited by the implicit assumptions that the entire community structure of a network has no hierarchy, and that a vertex can freely connect to any other vertex in the network. Consider the network in Fig.~\ref{fig:subcomm}(b). The two cliques are connected by a relatively large number of edges, which are unlikely due to chance. Therefore, the two cliques can be considered as a single community. On the other hand, it is evident that the edge density between the two cliques is much smaller than that within the cliques, indicating sub-structures within the community. In reality, the concept of communities may vary, depending on at what granularity the network is analyzed. For example, from the viewpoint of the General Secretary of the United Nation, each country may be a community, while from the viewpoint of an elementary school student, his definition of community may correspond to the classes in the school. 

It is imperative to note the intrinsic difference between the scenarios in Fig.~\ref{fig:subcomm}(a) and~\ref{fig:subcomm}(b). In Fig.~\ref{fig:subcomm}(a), the two sub-networks cannot be separated due to their small sizes relative to the entire network. Although the number of edges connecting the two components is higher than expected, the difference between the observed and expected number of inter-community edges is not statistically significant, i.e. the inter-community edges may have appeared just by chance. Therefore, we call the two sub-networks \emph{affiliated communities}. On the other hand, in Fig.~\ref{fig:subcomm}(b), the two sub-networks are statistically closely associated, which may indicate some functional relationships. Therefore, we call them \emph{associated communities}. Note that, however, there is no clear distinction between the two types of inter-community relationships.

\subsection{The HQcut algorithm}

In order to address the resolution limit problem of the modularity function, Fortunato and Barthelemy suggested a method that applies modularity optimization to each sub-network to identify sub-community structures~\cite{For2007}. Here we generalize the idea. We first apply \emph{Qcut} to obtain a community structure with the highest $Q$. We then apply \emph{Qcut} to each sub-network \emph{recursively}, while ignoring all the inter-community edges. A critical issue is, then, how to decide whether a community should be further partitioned or not.

Here we propose two criteria. First, if the modularity of partitioning a sub-network is below a threshold $minq$, it is an indication that the sub-network has no strong sub-community structure, and therefore should not be partitioned. Second, it has been shown that a network may have a high modularity by chance, especially if the network is sparse~\cite{Gui2004}. To overcome this problem, we estimate the statistical significance of the modularity using a Monte-Carlo method. For each sub-network, we apply \emph{Qcut} to obtain a modularity $q$. The sub-network is also randomly rewired with a procedure described in~\cite{MilRewire2003} to obtain $n$ random sub-networks, where each vertex has the same degree as in the original sub-network. The \emph{Qcut} algorithm is then applied to each random sub-network. We compute the statistical significance of $q$ using a \emph{Z}-score:
\begin{equation}
Z = \frac{q - \langle q \rangle} { \sigma_q}, 
\end{equation}
where $\langle q \rangle$ and $\sigma_q$ are the mean and standard deviation of the modularity values of the random sub-networks. A high $Z$-score indicates a statistically significant modularity of the sub-network, and therefore may correspond to real sub-community structures.

Most real-world networks have $Q \ge 0.3$~\cite{NewmanSIAM03}. Therefore, we use this value as the default value of $minq$. Second, we use a $Z$-score cutoff, $minz \ge 2$, which corresponds to a $p$-value of 0.05. As shown in the Supplemental file, the results are generally insensitive with respect to a wide range of parameter values.

\subsection{Differentiate affiliated and associated communities}
As we have discussed, both affiliated and associated sub-communities are non-separable by simply optimizing modularity. \emph{HQcut} can identify both types of sub-communities, but is unable to differentiate them. Therefore, after obtaining the result of \emph{HQcut}, we need to determine whether a pair of communities are associated or affiliated. For this purpose, we first identify pairs of communities whose merge would increase the modularity of the entire network. Then for each candidate community pair ($c_i$, $c_j$) connected by $e_{ij}$ edges, we use a Monte-Carlo method to estimate the probability that we would see at least $e_{ij}$ edges between them if the entire network were randomly rewired. We use the same rewiring procedure mentioned earlier~\cite{MilRewire2003}. We consider two communities as associated if the probability is smaller than 0.01, and affiliated if the probability is greater than 0.1. Those with intermediate probabilities are ignored, since we do not have enough statistical evidence about their relationships. Furthermore, we define a community as \emph{associated} if it is associated with another community, or \emph{affiliated} otherwise.

\section{Results}
In order to test the performance of our algorithms, we applied them to a variety of synthetic or real-world networks, and compared them with the Newman's algorithm (\emph{Newman})~\cite{Newman2006} as well as the simulated annealing algorithm (\emph{SA})~\cite{Gui2005}. The implementations of \emph{Newman} and \emph{SA} were obtained from the original authors. 

\subsection{Computer-generated networks}

We first considered three sets of computer-generated networks with known community structures, and compared the accuracy of the algorithms in identifying the known communities. Each network in these tests has 1000 vertices. 

The first set of networks was constructed as follows. The vertices in each network were divided into 20 communities of equal sizes. Edges were randomly placed between the vertices in the same community with a probability $p_{in}$, and across communities with a probability $p_{out}$. We chose $p_{in} = 0.3$, which corresponds to 15 intra-community edges for each vertex on average, and varied $p_{out}$ from 0.006 to 0.06. Note that although $p_{in} > p_{out}$, a vertex may have more inter-community edges than intra-community ones. 

The second set of networks differs from the first set in that the communities may have different sizes. To be precise, each network contains one community with 100 vertices, three communities with 40 vertices each, nine communities with 20 vertices each, and 40 communities with 15 vertices each. Edge probabilities were chosen such that each vertex has on average $n_{in} = 6 + \ln S$ intra-community edges, where $S$ is the size of the community that the vertex resides, and $n_{out} = 2$ to 24 inter-community edges, with an increment of two.

We designed a third set of networks to contain hierarchical communities. The vertices in each network were first grouped into ten equal-sized communities. Each community was then divided into two sub-communities. Edges were placed randomly with probability $p_{out} = 0.01$ between vertices in different communities, $p_1$ = 0.3 between vertices within the same sub-community, and $p_2 = 0.05$ between vertices within the same community but in different sub-communities. 

To measure the accuracy of a predicted community structure, we computed the Jaccard Index, which is based on the number of correctly identified intra-community vertex pairs~\cite{JaccardIndex}. Given a true community structure, $C1$, and a predicted community structure, $C2$, let $S1$ be the set of vertex pairs in the same community of $C1$, and $S2$ the set of vertex pairs in the same community of $C2$. The Jaccard Index is defined as
\begin{equation}
JI (S1, S2) = \frac{|S1 \cap S2|} {|S1 \cup S2| }.
\end{equation}
The value of Jaccard Index is in [0, 1], with one being the most accurate. The results using two other accuracy measurements, the Fowlkes-Mallows Index~\cite{Fow83} and Variation of Information~\cite{Mei2007}, are provided in the supplemental file.

As shown in Fig.~\ref{fig:res:qcut} (a)-(c), for the first set of networks, \emph{Qcut} and \emph{SA} clearly outperformed \emph{Newman} in optimizing $Q$. Furthermore, the slightly improved $Q$ values resulted in significantly better accuracies of community structures. In addition, when $p_{out}$ is small, \emph{Qcut} and \emph{HQcut} have almost the same results, indicating that \emph{HQcut} did not over-partition the communities. For large $p_{out}$ values, \emph{HQcut} has slightly lower modularity but higher accuracy than \emph{Qcut}. Indeed, for these networks, because of the abundance of inter-community edges, some communities were merged by \emph{Qcut} due to the resolution limit.

Fig.~\ref{fig:res:qcut}(d)-(f) show the results for the second set of networks. \emph{HQcut} was run with the default parameters, while the results are robust with respect to a wide range of parameter values (Supplemental file). As shown, \emph{Qcut} and \emph{SA} again found better modularities than \emph{Newman}. However, it is clear that for these networks, the higher modularities did not result in better community accuracies. In fact, the modularities found by \emph{Qcut} or \emph{SA} are often higher than those of the true community structures (Fig.~\ref{fig:res:qcut}(d)). \emph{HQcut}, on the other hand, have achieved the highest accuracy for all the networks, despite of slightly lower modularities. 

For small $n_{out}$ values, \emph{Newman} reached slightly better accuracies than \emph{Qcut} and \emph{SA} (Fig.~\ref{fig:res:qcut}(e)). However, the low accuracy of \emph{Qcut} was primarily caused by the merge of small communities, which can be easily resolved by a recursive algorithm such as \emph{HQcut}. In other words, by optimizing $Q$, we have a better chance to group together pairs of vertices that belong to the same community. In contrast, \emph{Newman} not only merged some small communities, but also assigned many vertices to wrong communities, which can not be resolved easily (Supplemental file).

For the third set of networks, \emph{Qcut} and \emph{Newman} successfully identified all communities with 100\% accuracy, but could not separate the sub-communities. In contrast, \emph{HQcut} successfully detected all sub-communities, with an accuracy of 99.9\%. Furthermore, with the statistical test proposed in this paper, we can distinguish the inter-community relationships in the second and the third sets of networks. The communities in the second set of networks are rarely associated with statistical significance, while  pairs of sub-communities in the third set of networks are often statistically significantly associated, indicating the existence of hierarchical communities (Supplementary file).

\subsection{Real-world networks}

As a further test of our algorithms, we applied them to several real-world networks, which may have different topological properties than the computer-generated networks. 

In the first real-world network, each vertex is a football team in the United States NCAA division I-A, and an edge between two teams represents a regular-season game played by them in year 2006. This network is interesting because of its known community structure. The 115 teams have been organized into eleven conferences (excluding the teams in the independence conference), and games were played more frequently between teams in the same conference than teams in different conferences. Therefore, each conference can be considered as a community. 

\begin{figure}[tbp]
\center 
\includegraphics[scale=.6]{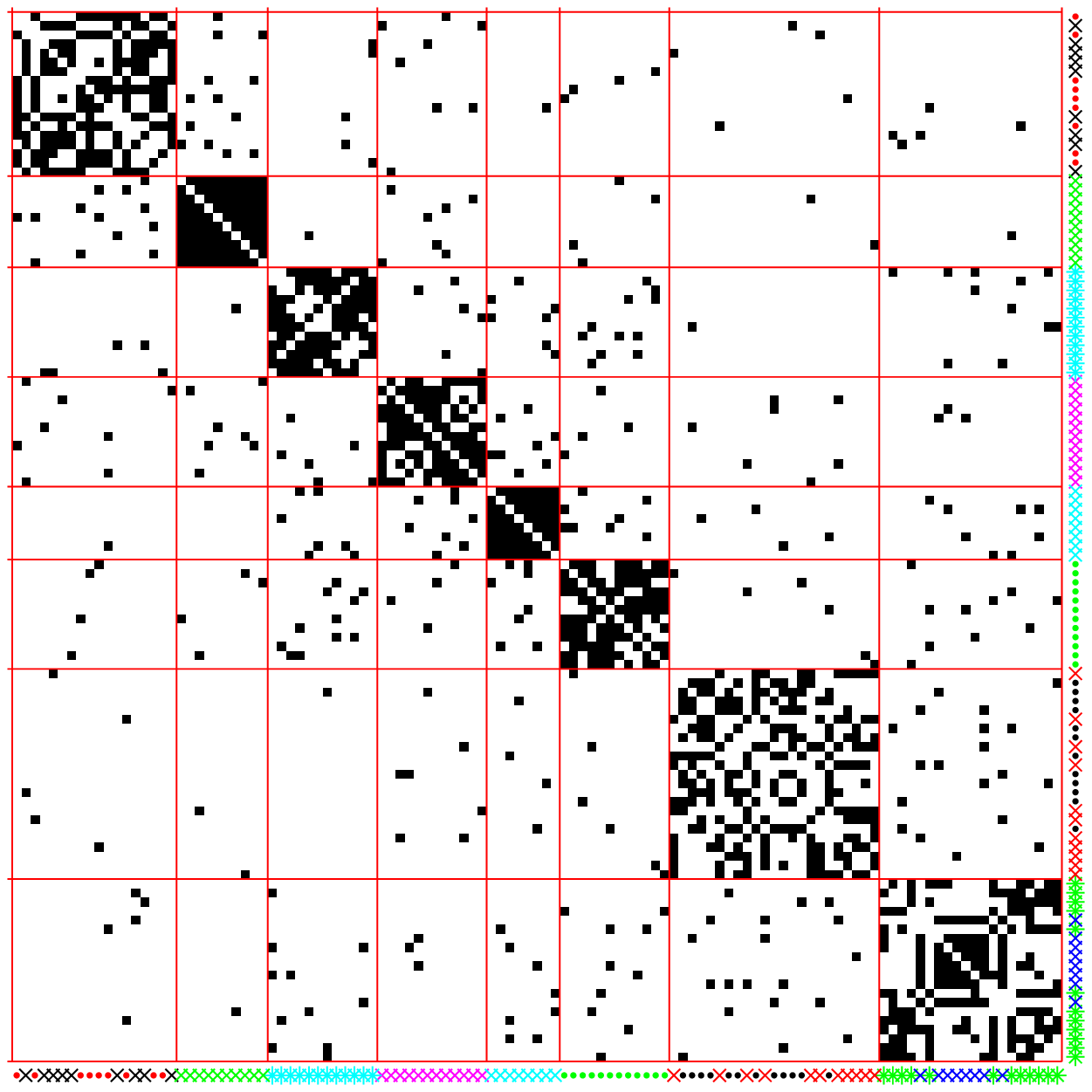} \hspace{.1in}
\includegraphics[scale=.6]{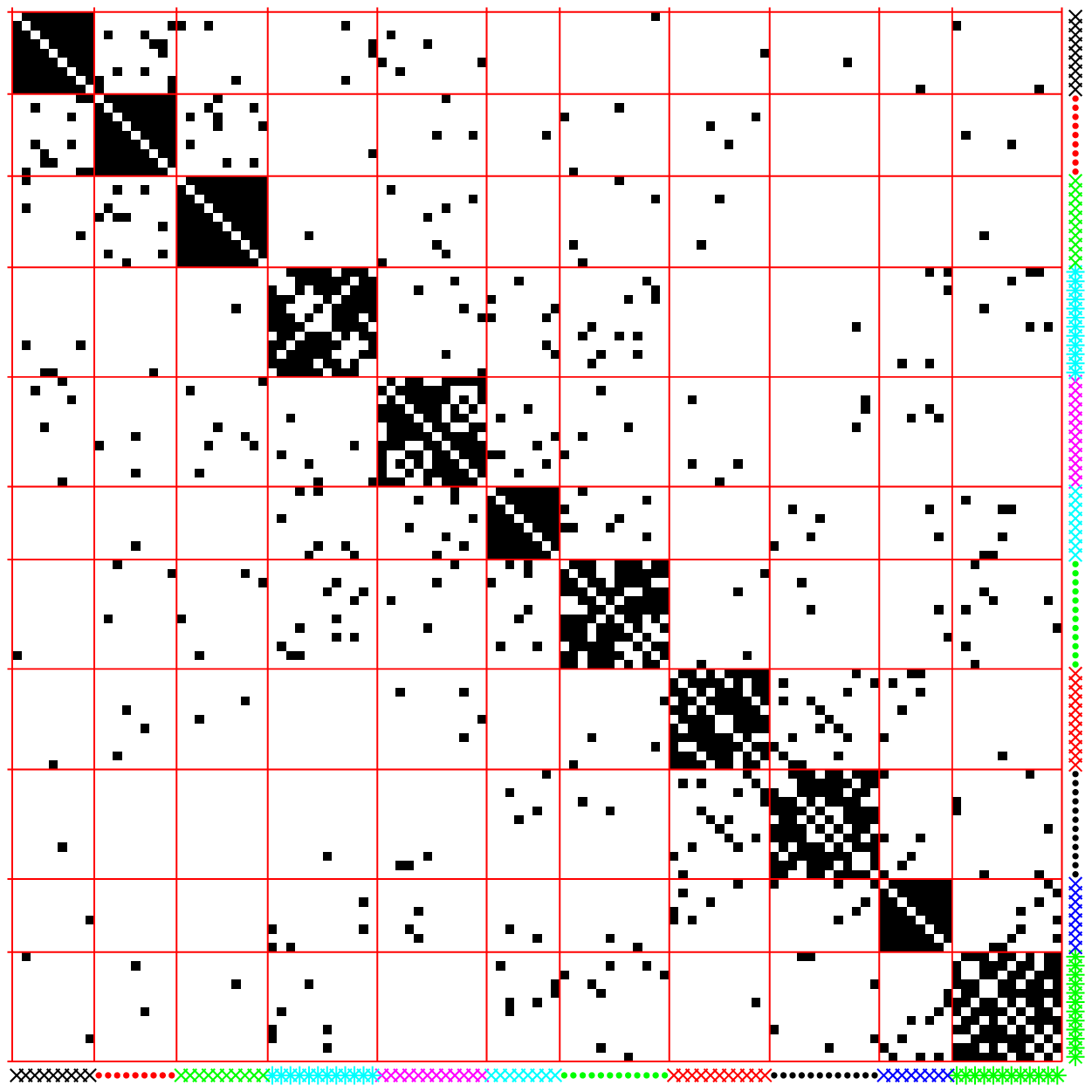}
\caption{{\label{fig:football}} Community structure in the football team network. Each symbol along the axes represents a conference. Left: results of \emph{Qcut}. Right: results of \emph{HQcut}. The 11 conferences shown in the right panel, from top to bottom, are WAC, Mountain West, Pacific-10, Conference USA, Big 12, Sun Belt, SEC, Big Ten, Mid-American, Big East, and ACC.
}
\end{figure}

Applying \emph{Qcut} to the network, we discovered eight communities ($Q = 0.608$), five of which matched individual conferences precisely (Pacific-10, Conference USA, Big 12, Sun Belt, and SEC) (Fig.~\ref{fig:football}). Each of the other three communities contains two conferences: one community contains WAC and Mountain West, one contains Big Ten and Mid-American, and the other contains Big East and ACC. The teams in these conferences have a relatively high frequency of inter-conference games with the teams in a conference that are geographically close. \emph{Newman} returned the same results as \emph{Qcut}. In contrast, with \emph{HQcut}, the network was divided into eleven communities ($Q = 0.596$), each of which corresponds to a conference precisely (Fig.~\ref{fig:football}). 

%The second example is a network of politics blogs~\cite{Newman2006}. The vertcies in the network correspond to 1222 blogs, and an edge connect two vertices if there is a hyper link from one blog to the other. \emph{Qcut} identified two large communities that cover all except eight blogs ($Q=0.4260$). The community structure identified by \emph{Newman} is almost identical, with slightly lower modularity ($Q=0.4257$). As expected, one community contains primarily conservative blogs (611/650, 94\%), while the other includes primarily liberal blogs (546/564, 96.8\%). Applying \emph{HQcut} to the network did not change the community structure, indicating that each community is indivisible.

We also tested the algorithms on a number of real-world networks with unknown community structures. For these tests, we were unable to measure the accuracy of the algorithms, due to the lack of known community structures. Therefore, we focused on the modularity values. As we have shown on the synthetic networks, although a higher modularity may not necessarily guarantee a better accuracy in community discovery, it nevertheless generally means better accuracy in recovering the true intra-community vertex pairs, which is necessary for a recursive algorithm such as \emph{HQcut} to succeed. 

The results on these networks are shown in Table~\ref{tab1}. The detailed information of the networks is included in the supplemental file. As shown, \emph{Qcut} always obtained higher modularities than \emph{Newman}. While \emph{SA} can achieve higher modularity for small networks, its performance on large networks is often worse than \emph{Qcut} and \emph{Newman}, even with much longer running time. The \emph{Newman} algorithm is faster than \emph{Qcut} on networks up to $\sim$1500 vertices, but slower than \emph{Qcut} for larger networks. 
 
Next, we applied \emph{HQcut} to these networks and compared the results to those in~\cite{For2007} (\emph{SA-2}), which were obtained by applying \emph{SA} to each community while ignoring the inter-community edges. Although \emph{SA-2} only allowed one level of hierarchy while \emph{HQcut} supported multiple levels of hierarchy, the latter usually returned fewer sub-communities than the former, indicating that \emph{SA-2} had probably over-partitioned these networks.

In order to test what type of communities are more abundant in the networks, we counted for each network the number of associated or affiliated communities as defined early. Interestingly, as shown in Table~\ref{tab1}, some networks consist of primarily affiliated communities  while other networks contain many associated communities, indicating hierarchical community structures in the latter group of networks. This preference seems to be unrelated to the edge density or the modularity of the networks and may deserve further studies. A possible explanation is that the edges in the latter group of networks (e.g. Circuit, PPI, and Internet) represent physical interactions. As a result, the interactions are limited by some spatial or structural constraints, and therefore a hierarchical community structure may be more feasible. In contrast, the edges in the former group represent some logical relationships, and therefore are not limited by such constraints. 

\subsection{An application to a biological network}

Finally, as a real application, we applied \emph{Newman}, \emph{Qcut}, and \emph{HQcut} to a protein-protein interaction network and studied the discovered communities in more detail. The network contains 2708 proteins and 7123 pairwise physical interactions in the yeast \emph{Saccharomyces cerevisiae}~\cite{Kro2006}. \emph{Newman} and \emph{Qcut} identified 56 ($Q = 0.694$) and 93 communities ($Q = 0.696$), respectively, while \emph{HQcut} detected 316 communities ($Q = 0.582$). In order to determine the biological significance of the communities, we compared the communities to the known protein complexes in the MIPS database~\cite{Mew2006}. Note that protein complexes in the MIPS database are also organized into some hierarchical structures, i.e. a large protein complex may contain several smaller complexes. A protein may also belong to multiple protein complexes. In order to measure how well a discovered protein community represents real protein complex, we computed a matching score for a community $c$ as follows.
\begin{equation}
Matching(c) = \max_i |c \cap p_i| / \sqrt{|c \cap P| \times |p_i \cap C|},
\end{equation}
where $p_i$ is the $i$-th known protein complex, and $c \cap p_i$ is the set of proteins shared between $c$ and $p_i$. $C$ and $P$ represent the set of all proteins in the network or in the MIPS protein complex database, respectively. The overall performance of the algorithm was measured by the weighted average of matching scores for all communities.

\begin{figure}[tbp]
\center 
\includegraphics[scale=1.5]{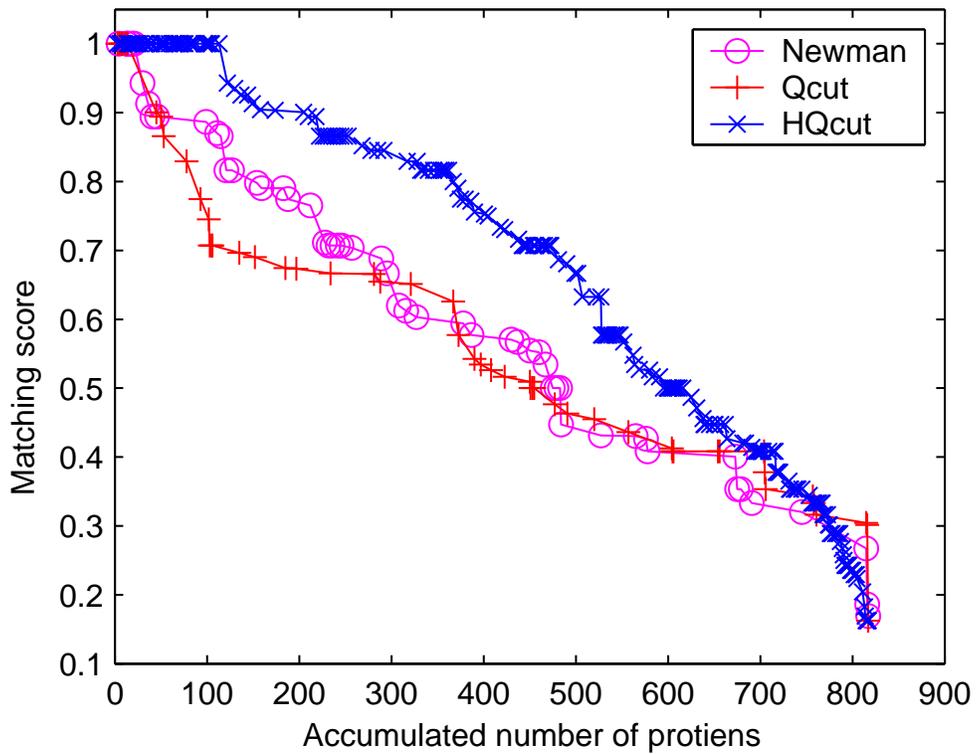}
\caption{{\label{fig:ppiMatching}} Matching score for protein communities. Each point represents a community, sorted according to their matching scores. The x-axis shows the accumulated number of proteins in the communities exceeding a given matching score on the y-axis.
}
\end{figure}

Fig.~\ref{fig:ppiMatching} shows the total number of proteins as a function of matching scores. Overall, 216 communities identified by \emph{HQcut} matched to some known complexes, with an weighted average matching score of 0.70. In comparison, 52 communities by \emph{Qcut} and 53 by \emph{Newman} matched to some known complexes, with average matching scores of 0.55 and 0.56, respectively. Furthermore, \emph{HQcut} discovered 43 communities that perfectly matched to some known protein complexes. In contrast, \emph{Qcut} found seven perfectly matched communities, and \emph{Newman} only identified five such communities. Therefore, by allowing sub-communities, \emph{HQcut} recovered a large number of real protein complexes, while each community identified by \emph{Newman} or \emph{Qcut} may contain several protein complexes.

 %Consequently, the Newman's algorithm has slightly better matches with the known protein complexes, similar to what we have seen in the computer-generated networks. 

\begin{figure}[tbp]
\center 
\includegraphics[scale=.7]{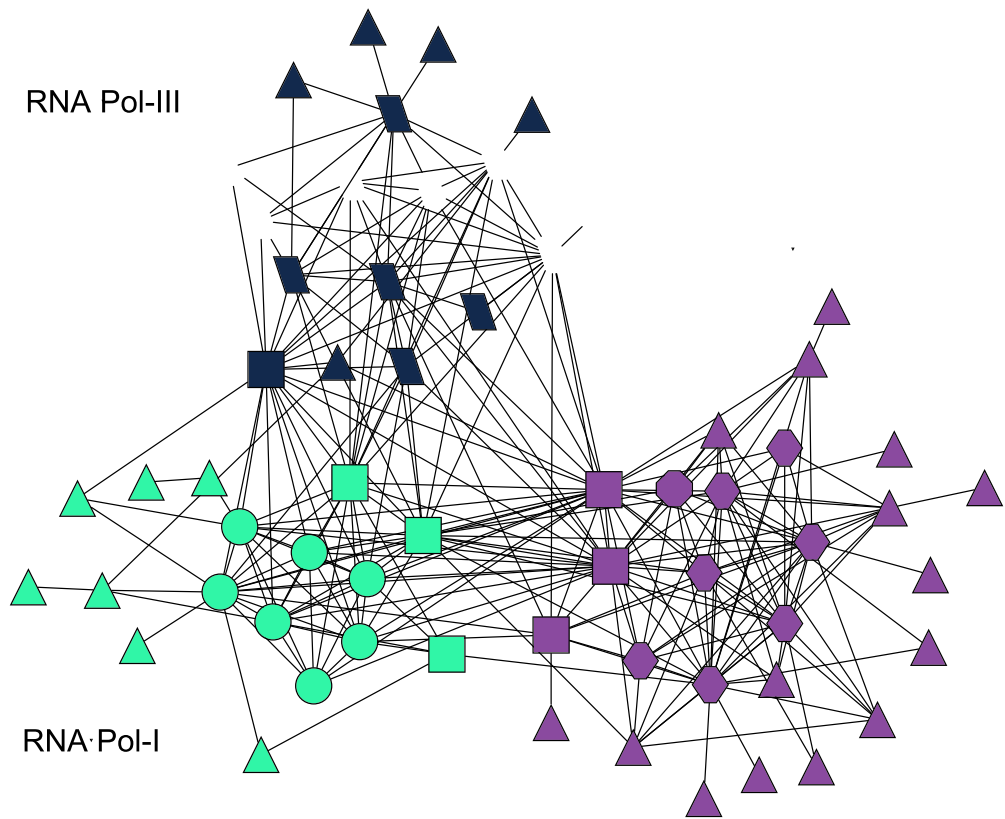}
\caption{{\label{fig:ppiExamp}} Example hierarchical communities in PPI network. The three colors represent three sub-communities discovered by \emph{HQcut} in a community identified by \emph{Qcut}. Circles, hexagons, and parallelograms are known components of RNA polymerase I, II, and III, respectively. Squares are shared components of two or three RNA polymerases. Triangles are proteins that are not components of the three complexes by current knowledge.}
\end{figure}

We again computed the numbers of affiliated and associated communities in this network, and found that it contains more associated communities than affiliated ones (195 vs. 83), indicating that the majority of the additional communities found by \emph{HQcut} are due to hierarchical communities. To analyze whether the hierarchical structures have any biological significance, we manually inspected the associated communities. Interestingly, almost all of the statistically significantly associated communities are biologically related. For example, the three RNA polymerases, RNA pol I, RNA pol II, and RNA pol III were identified as a single community by \emph{Qcut}, but were further partitioned by \emph{HQcut} into three sub-communities (Fig.~\ref{fig:ppiExamp}). The three communities are highly associated due to a few common components shared by the three complexes. In another example, snRNA sub-units U1, U2 and U6 were also identified as a single community by \emph{Qcut} but separated into three sub-communities by \emph{HQcut}. Other examples include SAGA and TFIID complexes, INO80 and SWR1 complexes, as well as eIF-2B and eIF-3 complexes (Supplemental file). Therefore, by combining the results of \emph{Qcut} and \emph{HQcut}, we are able to reveal the true hierarchical community structures of the network.

\section{Conclusions and discussion}

In this paper, we described an efficient algorithm, \emph{Qcut}, for discovering communities from complex networks by optimizing the modularity function. We showed that the algorithm can find a higher modularity than the existing algorithms on both computer-generated and real-world networks. When the communities are not so small and the inter-community connectivities are sparse, a higher modularity indeed represents a better community discovery accuracy. On the other hand, we also showed that, when a network contains small or hierarchical communities, optimizing modularity may fail to reveal the fine structures at a satisfactory resolution. To circumvent this problem, we proposed a recursive algorithm, \emph{HQcut}, which provides a higher resolution without introducing spurious communities. Using a variety of synthetic as well as real-world networks with known community structures, we demonstrated that \emph{HQcut} can achieve a much higher accuracy than algorithms based on modularity optimization alone. We also studied a protein-protein interaction network, and found that the protein communities identified by \emph{HQcut} correspond to known protein complexes very well, while each community found by modularity optimization may contain several protein complexes.

Our results may first seem to suggest that modularity optimization is not a very good strategy for community discovery in practice. Nevertheless, the success of \emph{HQcut} largely depends on the effectiveness of \emph{Qcut} to optimize $Q$. By optimizing modularity, the \emph{Qcut} algorithm may merge several communities into a single one, which can be easily separated by a recursive algorithm such as \emph{HQcut} in this paper. In contrast, algorithms that did not succeed in optimizing modularity may split the members of a community into several communities, a mistake that cannot be easily recovered in a post-processing way.

Finally, we proposed a statistical significance test to differentiate the two scenarios that may cause the resolution limit: small communities or hierarchical communities. By combining \emph{HQcut} with the significance test, we are able to not only detect communities with a high resolution, but also identify pairs of highly associated communities. As shown in the case of protein-protein interaction networks, these statistically associated community pairs are indeed functionally related, and form a community at a higher hierarchy. Since many real-world networks are hierarchical, identifying and analyzing such structures will be an essential step towards understanding their organizing principles in general.

\section*{Acknowledgments}
This research was supported in part by NSF grants
ITR/EIA-0113618 and IIS-0535257, and a grant from Monsanto Company. The authors wish to thank Mark E. J. Newman, Roger Guimera and Luis A.N. Amaral for sharing their programs, and Uri Alon for sharing network data.

\bibliographystyle{plain}
\bibliography{netQ,supplemental}

\clearpage

\begin{table*}
\center{
\caption{\label{tab1} Community results on real-world networks.}
\footnotesize{
\begin{tabular}[tbp]{lrr|rrr|rrr|rrr|rr|rrr}\hline\hline
	\multicolumn{3}{c|}{	Network	}&		\multicolumn{3}{|c|}{\textit{Newman}}		&		\multicolumn{3}{|c|}{\textit{SA}}	&		\multicolumn{3}{|c|}{\textit{Qcut}}	&		\multicolumn{2}{|c|}{\textit{SA-2}}	&		\multicolumn{3}{|c}{\textit{HQcut}}	\\\hline
Name	&	$N$	&	$M$	&	$k$	&	$Q$	&	Time	&	$k$	&	$Q$	&	Time	&	$k$	&	$Q$	&	Time	&	$k$	&	$Q$	&	$k$	&	$Q$	&		Af/As		\\\hline
Social		&		67		&		142		&		8	&	0.573		&	0.01	&		10	&	0.608		&	 5.4 	&		8	&	0.587		&	2	&	21	&	0.532	&	9	&	0.578	&		9/0		\\
Neuron		&		297		&		2359		&		4	&	0.396		&	0.4	&		4	&	0.408		&	 139 	&		4	&	0.398		&	1.9	&	20	&	0.319	&	10	&	0.363	&		2/6		\\
Ecoli Reg		&		418		&		519		&		38	&	0.766		&	0.7	&		27	&	0.752		&	 147 	&		39	&	0.776		&	12.7	&	76	&	0.661	&	44	&	0.769	&		40/0		\\
Circuit		&		512		&		819		&		15	&	0.804		&	1.8	&		11	&	0.670		&	 143 	&		13	&	0.815		&	6.1	&	70	&	0.64	&	43	&	0.723	&		9/15		\\
Yeast Reg		&		688		&		1079		&		26	&	0.759		&	3	&		9	&	0.740		&	 22.5m 	&		27	&	0.766		&	13.4	&	57	&	0.677	&	66	&	0.696	&		28/13		\\
Ecoli Met		&		563		&		709		&		29	&	0.827		&	2.06	&		19	&	0.828		&	200.4	&		21	&	0.835		&	12	&		92		&	0.728	&	37	&	0.81	&		21/2		\\
Ecoli PPI		&		1440		&		5871		&		18	&	0.367		&	33.2	&		14	&	0.387		&	 97.8m 	&		21	&	0.387		&	41.5	&		88		&	0.305	&	112	&	0.346	&		10/31		\\
Internet 	&	3015	&	5156	&	46	&	0.611	&	253.7	&	20	&	0.624	&	 184m 	&	21	&	0.634	&	43	&	219 	&	0.556	&	186	&	0.566	&	 32/59 	\\
\hline\hline
\end{tabular}
\parbox[]{7in}{The results by \emph{SA} and \emph{SA-2} on the first five networks were directly obtained from~\cite{For2007}. The unit of time is second, unless m (minute) was specified. The last column shows the numbers of affiliated (Af) versus associated (As) communities.}
}
}
\end{table*}

\begin{figure*}[tbp]
\center 
\includegraphics[width=.45\textwidth]{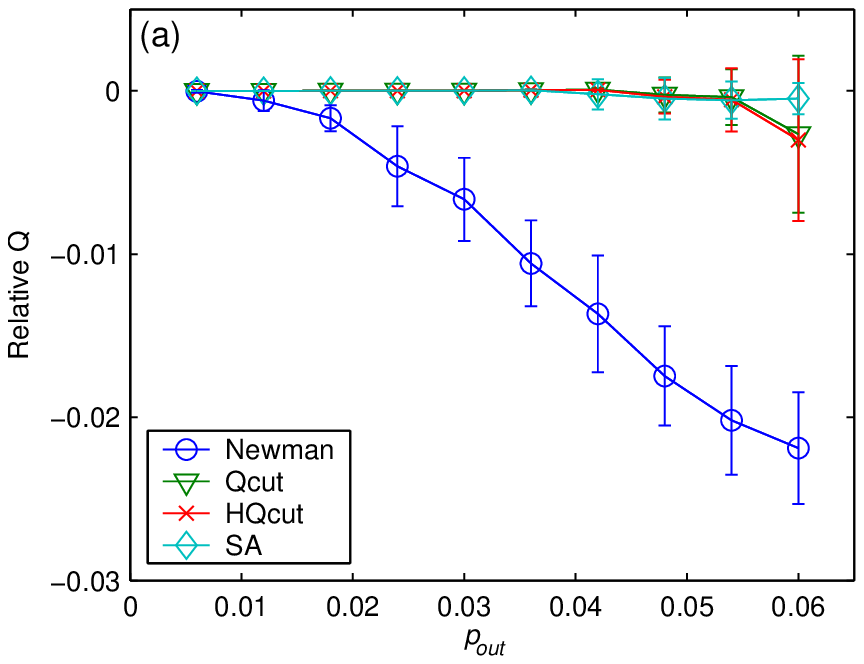}
\includegraphics[width=.45\textwidth]{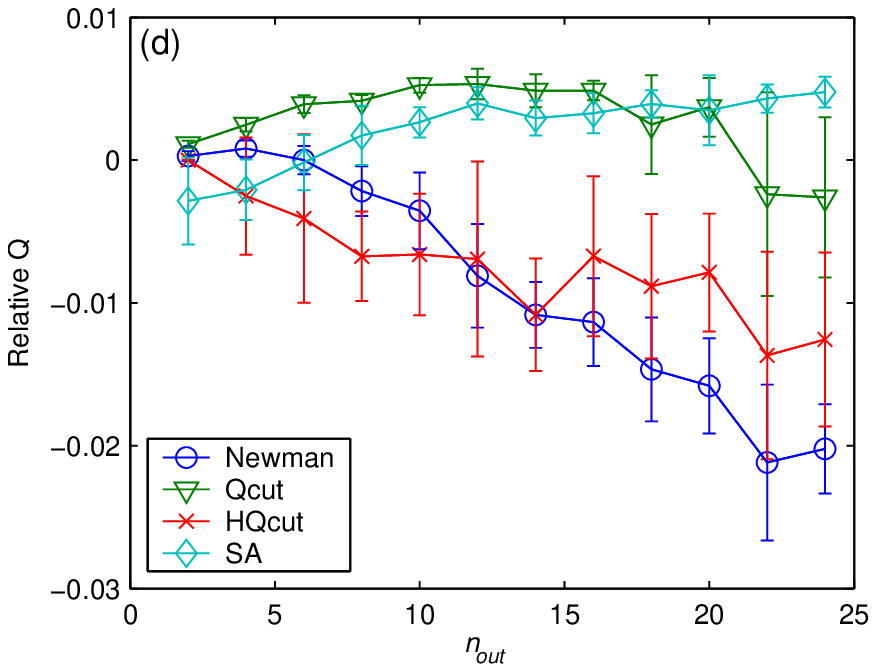}\\
\includegraphics[width=.45\textwidth]{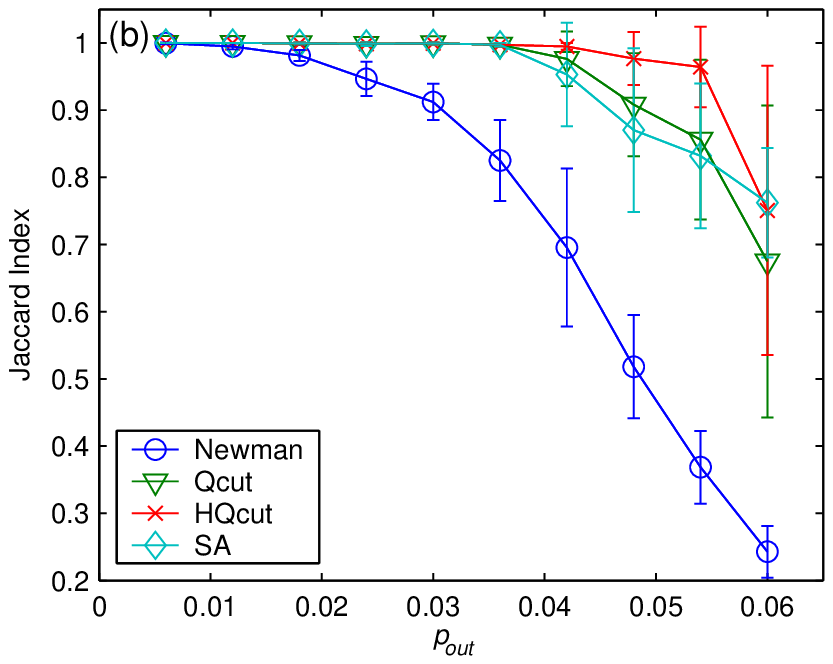}
\includegraphics[width=.45\textwidth]{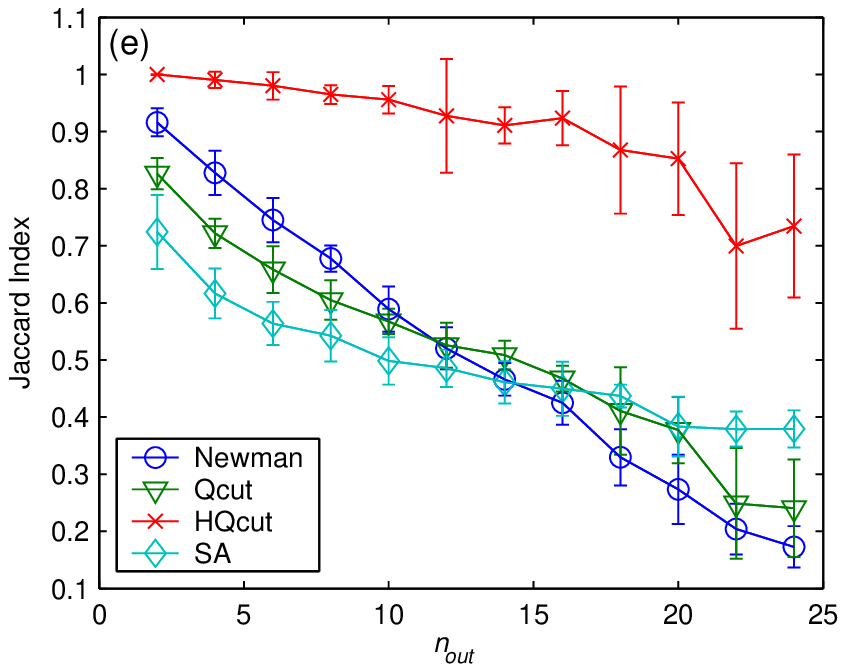}\\
\includegraphics[width=.45\textwidth]{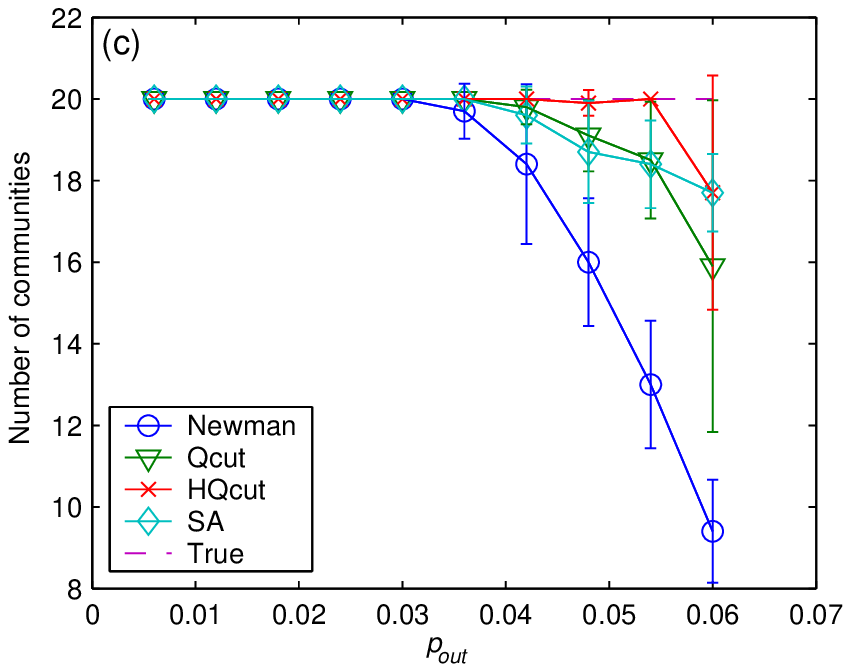}
\includegraphics[width=.45\textwidth]{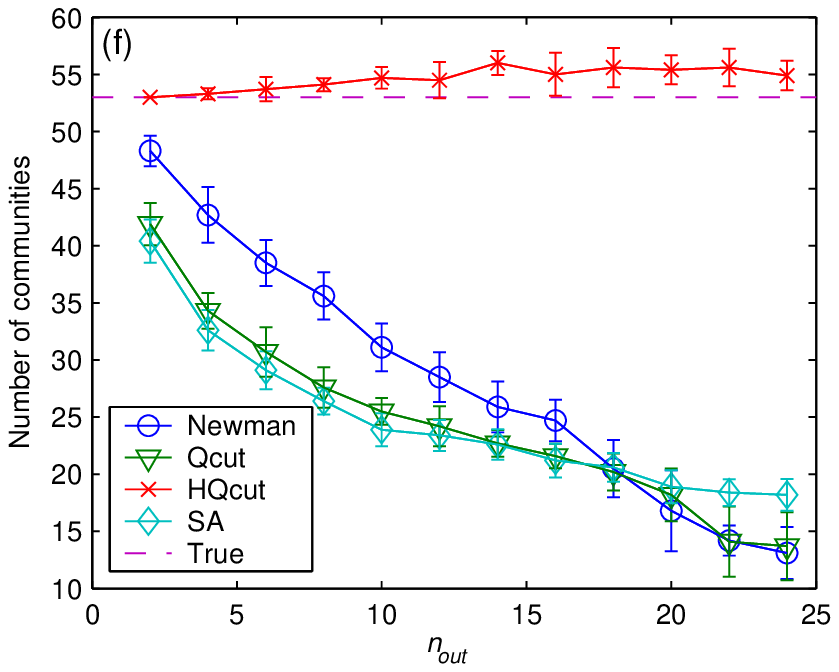}
\caption{{\label{fig:res:qcut}} Results on computer-generated networks. Relative $Q$ = $Q_{found} - Q_{true}$. (a)-(c): networks with equal community sizes. (d)-(f): networks with diverse community sizes. Each data point is the mean and standard deviation of 100 networks. }
\end{figure*}

\end{document}